\documentstyle[twocolumn,psfig,epsfig]{mn}

\newif\ifAMStwofonts
\AMStwofontstrue
\def\ltsim{\lower.5ex\hbox{$\; \buildrel < \over \sim \;$}}
\def\gtsim{\lower.5ex\hbox{$\; \buildrel > \over \sim \;$}}
\def\ltsim{\lower.5ex\hbox{$\; \buildrel < \over \sim \;$}}
\def\gtsim{\lower.5ex\hbox{$\; \buildrel > \over \sim \;$}}

\newcommand{\ignore}[1]{}
\newcommand{\beq}[1]{\begin{equation}\label{#1}}
\newcommand{\eeq}{\end{equation}}
\newcommand{\beqa}[1]{\begin{equation}\label{#1}\begin{eqalign}}
\newcommand{\eeqa}{\end{eqalign}\end{equation}}
\newcommand{\bsubeq}[1]{\begin{subequations}\label{#1}\begin{eqalignno}}
\newcommand{\esubeq}{\end{eqalignno}\end{subequations}}

\def \G   {{ G}}
\def \g   {{ g}} 

\def \C   {{ C}}
\def \c   {{ c}}

\def \Lm  {{ \Lambda}}
\def \l {{ \lambda}}

\def \L  {{\cal L}}
\def \bb  {{\beta^{-1}}}

\begin{document}
\title{Objective Classification of Galaxy Spectra using the Information Bottleneck Method}
\author[Slonim et al. ] {Noam Slonim$^1$, Rachel Somerville$^{2,3}$,
Naftali Tishby$^{1}$ and  Ofer Lahav$^{2,3}$\\
$^1$ Institute of Computer Science and Center for Neural Computation, 
The Hebrew University, Jerusalem 91904, Israel\\
$^2$ Institute of Astronomy, Madingley Rd., CB3 0HA, Cambridge, UK \\
$^3$ Racah Institute of Physics, The Hebrew University, Jerusalem 91904, Israel\\
}

\maketitle

\begin{abstract}
A new method for classification of galaxy spectra is presented, based
on a recently introduced information theoretical principle, the {\em
information bottleneck}. For any desired number of classes, galaxies
are classified such that the information content about the spectra is
maximally preserved.  The result is classes of galaxies with similar
spectra, where the similarity is determined via a measure of
information. We apply our method to $\sim 6000$ galaxy spectra from the
ongoing 2dF redshift survey, and a mock-2dF catalogue produced by a
Cold Dark Matter-based semi-analytic model of galaxy formation. We find a good
match between the mean spectra of the classes found in the data and in
the models.  For the mock catalogue, we find that the classes produced
by our algorithm form an intuitively sensible sequence in terms of
physical properties such as colour, star formation activity,
morphology, and internal velocity dispersion. We also show the correlation
of the classes with the projections resulting from a Principal
Component Analysis.
\end{abstract}

\section{Introduction} 
Very large numbers of galaxy spectra are being generated by modern
redshift surveys: for example, the Anglo-Australian Observatory
2-degree-Field (2dF) Galaxy Survey (e.g. Colless 1998; 
Folkes et al. 1999) aims to collect
250,000 spectra and has already gathered over $100,000$ redshifts (as
of May 2000).  The Sloan Digital Sky Survey (e.g. Gunn \& Weinberg
1995; Fukugita 1998) will observe the spectra of millions of
galaxies. Such large data sets will provide a wealth of information
pertaining to the distribution and properties for a vast variety of
galaxy types. A major goal of such surveys is to determine the
relative numbers of these different galaxy populations, and eventually
to gain clues as to their physical origin. Traditional methods of
classifying galaxies ``by eye'' are clearly impractical in this
context. The analysis and full exploitation of such data sets require
well justified, automated and objective techniques to extract as much
information as possible. In this paper we present a new approach to
objective classification of galaxy spectra, by utilising a recently
proposed method based on information theory.

Broadly speaking, inference from galaxy spectra can be considered in
two ways.  One approach is to compare each galaxy spectrum to the most
likely one of a library of model spectra (e.g. based on age,
metallicity and star formation history). The other model independent
approach is to consider an ensemble of observed spectra and to look
for patterns in analogy with the stellar HR-diagram or the Hubble
sequence for galaxy morphology.  

The concept of spectral classification goes back to 
Humason (1936) and Morgan \& Mayall (1957).
Recent attempts to analyse galaxy
properties from spectra in a model independent way have been made
using Principal Component Analysis (e.g. Connolly et al. 1995; Folkes,
Lahav \& Maddox 1996). The PCA is effective for data compression, but
if one wishes to break the ensemble into classes it requires a further
step based on a training set (e.g. Bromley et al. 1998, Folkes et
al. 1999).  Unlike PCA, the {\em information bottleneck (IB) method}
presented here is non-linear, and it naturally yields a principled
partitioning of the galaxies into classes.  These classes are obtained
such that they maximally preserve the original information between the
galaxies and their spectra.

The end goal of galaxy classification is a better understanding of the
physical origin of different populations and how they relate to one
another. In order to interpret the results of any objective
classification algorithm, we must relate the derived classes to the
physical and observable galaxy properties that are intuitively
familiar to astronomers. For example, important properties in
determining the spectral characteristics of a galaxy are its mean
stellar age and metallicity, or more generally its full star formation
history. This is in turn presumably connected with
\emph{morphological} properties of the galaxy; eg. its Hubble or
T-type. An assumed star formation history can be translated into a synthetic
spectrum using models of stellar evolution (e.g., Bruzual \& Charlot
1993, 1996; Fioc \& Rocca-Volmerange 1997). Spectral features are also affected by
dust reddening and nebular emission lines. 

In one example, typical of such an approach 
(Ronen, Aragaon-Salamance \& Lahav 1999), the
star formation history was parameterized as a simple single burst or
an exponentially decreasing star formation rate. However, the
construction of the ensemble of galaxy spectra was done in an ad-hoc
manner.  Here we try a similar exercise using a \emph{cosmologically
motivated ensemble} of synthetic galaxies, with realistic star
formation histories. These histories are determined by the physical
processes of galaxy formation in the context of hierarchical structure
formation.  We construct such an ensemble using semi-analytic models
of galaxy formation set within the Cold Dark Matter (CDM) framework.
While this approach has the disadvantage of relying on numerous
assumptions about poorly-understood physics, it has the advantage of a
certain self-consistency, and of producing ensembles of galaxies with
global properties that agree well with observations (eg. local
luminosity functions, colours, Tully-Fisher relations,
metallicity-luminosity relation, etc).

In this paper, we shall analyze a subset of about $6000$ spectra from
the ongoing 2dF survey. Using the semi-analytic models we construct a
``mock-2dF'' catalogue with the same magnitude limit, spectral
resolution, wavelength coverage and noise as in the 2dF data.  We
analyze the synthetic and observed spectra in the same way using the
IB method. Using the mock catalogue, we then interpret the resulting
classes in terms of familiar, intuitive physical properties of
galaxies. We also draw a connection between the IB approach and the
results of a PCA analysis for both the 2dF and mock data.

The organization of the rest of this paper is as follows. In section 2
we present the IB method and the classification algorithm.  In section 3
we describe the sample of observed spectra and the semi-analytic
models used to produce the mock galaxy catalogues. In section 4 we
present our results, and in section 5 we relate the IB classification
to PCA.  We summarize our conclusions in section 6.

\section{The Classification Algorithm: the Information Bottleneck Method}

Consider a galaxy spectrum as an array of photon counts in different
wavelength bins. 
\footnote{
We note that other representations of the spectra are possible
(e.g. flux instead of counts).}  Thus each galaxy is represented in a
high-dimensional space, where each component corresponds to the counts
in a given spectral bin.  We can also view the ensemble of such
spectra as the joint distribution of the galaxy-wavelength variables.
By normalizing the total photon counts in each spectrum to unity, we
can consider it as a conditional probability, the probability of
observing a photon at a specific wavelength from a given galaxy.
This view of the ensemble of spectra as a conditional probability
distribution function enables us to undertake the information
theory-based approach that we describe in this section.

Our goal is to group the galaxies into classes that preserve some
objectively defined spectral properties.  Ideally, we would like to
make the number of classes as small as possible (i.e. to find the
`least complex' representation) with minimal loss of the `important'
or `relevant' information.  In order to do this objectively, we need
to define formal measures of `complexity' and `relevant information'.

Some classification methods are based on a training set of labeled
data (e.g. morphological types of galaxies defined by a human
expert; 
e.g. Naim et al. 1995a,b).
Such prior labels introduce a bias towards existing
classification schemes.  On the other hand, our goal here is to
develop an {\em unsupervised} classification method, which is free of
this bias, and thus to provide objective, `meaningful',
categorization.  This problem, however, is ill defined without a
better definition of `meaningful'.  Almost all existing algorithms
begin with some pairwise `distances' (e.g. Euclidean) between the
points in the high-dimensional representation space, or with a
distortion measure between the data points and candidate group
representative or model.  The `meaning' is then dictated through this,
sometimes arbitrary, choice of the distance or distortion measure.
In addition, it is difficult in such cases to objectively evaluate the
quality of the obtained classes.

Recently, Tishby, Pereira \& Bialek (1999) proposed an information
theoretical approach to this problem which avoids the arbitrary choice
of the distance or distortion measures. It also provides a natural
quality measure for the resulting classification. Their algorithm is
extremely general and may be applied to different problems in
analogous ways.  This method has been successfully applied to the
analysis of neural codes (Bialek, Nemenman \& Tishby 2000), linguistic data
(word sense-disambiguation, Pereira, Tishby \& Lee, 1993) and for
classification of text documents (Slonim \& Tishby 2000). In the
latter case for example, one may see an analogy between an ensemble of
galaxy spectra and a set of text documents. The words in a document
play a similar role to the wavelengths of photons in a galaxy
spectrum, i.e. the frequency of occurrence of a given word in a given
document is equivalent to the number of photon counts at a given
wavelength in a given galaxy spectrum. In both cases, the specific
patterns of these occurances may be used in order to classify the
galaxies or documents.

\subsection{The concept of mutual information}
In the following we denote the set of galaxies by $\G$ and the array
of wavelength bins by $\Lm$.  As already mentioned, we view the
ensemble of spectra as a joint distribution $p(\g,\l)$, which is the
joint probability of observing a photon from galaxy $\g \in \G$ at a
wavelength $\l \in \Lm$.  We normalize the total photon counts in each
spectrum (galaxy) to unity, i.e. we take the prior probability $p(\g)$
of observing a galaxy $\g$ to be uniform: $p(\g) =
\frac{1}{N_{G}}$, where $N_{G}$ is the number of galaxies in this
sample $\G$. This is a standard statistical procedure since these
galaxies are sample points from the underlying (unknown) galaxy
distribution. Note, however, that the prior probability $p(\l)$ of
observing a photon in a wavelength $\l$ is not considered uniform.

Given two random variables, a fundamental question is: to what extent
can one variable be predicted from knowledge of the other variable?
Clearly, when the two variables are statistically independent, no
information about one variable can be obtained through knowledge of
the other one. This question is quantitatively answered through the
notion of {\em mutual information} between the variables. Moreover,
this is the {\em only possible measure}, up to a multiplicative
constant, that captures our intuition about the information between
variables, as was shown by Shannon (1948a,b) in his seminal work (see
also Acze`l and Darotzky, 1975).  In our case, the two variables are
the galaxy identity and the photon counts as a function of wavelength
(spectral densities). These variables are clearly not independent, as
the galaxy identity determines its spectrum through its physical
properties.
 
The mutual information between two variables can be shown (see
e.g. Cover \& Thomas 1991) to be given by the amount of uncertainty in
one variable that is removed by the knowledge of the other one. In our
case this is the reduction of the uncertainty in the galaxy identity
through the knowledge of its spectrum.  The uncertainty of a random
variable is measured by its {\em entropy}, which for the case of the
galaxy variable $\G$ is given by
\begin{equation}
\label{entropy}
H(\G) = - \sum_{\g} p(\g) \log p(\g) ~.
\end{equation}
Since in our case $p(\g)$ is uniform (over the sample) 
we get $~ H(\G) = \log N_{G} ~$. 
\footnote{
When we take the logarithm to base $2$ the information
is measured in bits.
This means that in the absence of other knowledge, 
the amount of information needed to 
specify a galaxy $\g$ out of the sample $\G$
is exactly $\log N_{G}$ bits.}

The amount of uncertainty in the galaxy identity, given its spectral density, 
is given by the {\em conditional entropy} of the galaxies on their spectra.
More formally stated, the conditional entropy of $\G$ given $\Lm$ is defined by
\begin{equation}
\label{cond_entropy}
H(\G|\Lm) = - \sum_{\l} p(\l) \sum_{\g} p(\g|\l) \log p(\g|\l) ~.
\end{equation}
Obviously, knowledge about $\Lm$ can only 
reduce the uncertainty in $\G$, i.e.
$~ H(\G|\Lm) \leq H(\G) ~$.
The amount of reduction in the uncertainty is thus 
the mutual information, which is now given by
\begin{equation}
\label{MI1}
I(\G;\Lm) = H(\G) - H(\G|\Lm) = 
\sum_{\g,\l} p(\g,\l) \log \frac {p(\g,\l)}{p(\g)p(\l)} ~,
\end{equation}
or, using $p(\g,\l) = p(\g)p(\l|\g)$, 
\begin{equation}
\label{MI2}
I(\G;\Lm) = \sum_{\g,\l} p(\g) p(\l|\g) \log \frac {p(\l|\g)}{p(\l)} ~.
\end{equation}
It is easy to see that $I(\G;\Lm)$ is symmetric and non-negative, and
is equal to zero if and only if $\g$ and $\l$ are independent.  As
$I(\G;\Lm)$ measures the reduction of uncertainty in $\G$ for known
$\Lm$, it is a measure of the amount of information about the galaxy
identity contained in the spectrum.

\subsection{The Bottleneck Variational Principle} 

Our goal is to find a mapping of the galaxies $\g \in \G$ into classes
$c \in \C$ such that the class $c(g)$ provides essentially the same
prediction, or information, about the spectrum as the specific
knowledge of the galaxy. The partitioning may be {\em ``soft''},
i.e. each galaxy is associated with each class through the conditional
probability $p(\c|\g)$.

The prior probability for a specific class $\c$ is then given by
\begin{equation}
\label{r1}
p(\c) =\sum_{\g} p(\g) p(\c|\g) ~.
\end{equation}
Using the fact that the only statistical dependence of $\Lm$ on  $\C$ 
is through the original statistical dependence of $\Lm$ on $\G$ 
(since the distribution of $\C$ is determined completely by
$p(\c|\g)$, a Markov condition) we get,
\begin{equation}
\label{r2}
p(\l|\c) = \sum_{\g}p(\l|\g)p(\g|\c) ~,
\end{equation}
where $p(\l|\c)$ can be clearly interpreted as the spectral density
associated with the class $\c$.  Using these equations, we can now
calculate the mutual information between a set of {\em galaxy classes} $\C$
and the {\em spectral wavelengths} $\Lm$. Specifically, this information is
given by
\begin{equation}
\label{MI3}
I(\C;\Lm) = \sum_{\c,\l} p(\c) p(\l|\c) \log \frac {p(\l|\c)}{p(\l)} ~.
\end{equation}

A basic theorem in information theory, known as {\em data processing
inequality}, states that no manipulation of the data can increase the
amount of (mutual) information given in that data.  Specifically this
means that by grouping the galaxies into classes one can only {\em
lose} information about the spectra, i.e.  $I(\C;\Lm) \leq I(\G;\Lm)$.
Our goal is then to find a non-trivial classification of the galaxies
that preserves, as much as possible, the original information about
the spectra. In other words, we wish to maximize $I(C;\Lm)$.  However,
based on the above inequality, maximizing this information is trivial:
each galaxy $g$ is a class $\c$ of its own, which formally means $\C
\equiv \G$.  To avoid this trivial solution, one must introduce a
formal {\em constraint} that will force the classification into a more
compact representation.

It turns out that the compactness of a classification is directly 
governed by the mutual information between the {\em classes and 
the galaxies}. This mutual information is given by
\begin{equation}
\label{MI4}
I(\C;\G) = \sum_{\c,\g} p(\g) p(\c|\g) \log \frac {p(\c|\g)}{p(\c)} ~.
\end{equation}
To understand this expression, it is useful to consider its behaviour
at two extremes.  One extreme is when the new representation is the
most compact one possible, i.e. there is only one class and all $\g \in
\G$ are assigned to it with probability $1$.  In this case there is no
dependence between $\G$ and $\C$, thus $I(\C;\G)$ is trivially
minimized to zero.  This agrees with our intuition that a single
global class carries no information at all about the original identity
of a galaxy (i.e. its unique spectrum).  In the other extreme, the
classification is maximally complex when every $\g \in \G$ is assigned
to a class of its own, i.e. $\C \equiv \G$.  In this case $I(\C;\G)$
is maximized.  Accordingly, the class of a specific galaxy provides
the full information about its identity.  The interesting cases are of
course in between, where the number of classes is relatively small
(but larger than one). In fact, in general the mutual information
$I(\C; \G)$ gives a well justified measure for the complexity of the
classification (Tishby, Pereira
\& Bialek 1999). Moreover, the maximal amount of information that the
class can provide about the spectrum, $I(\C;\Lm)$, for a given amount
of information preserved about the galaxies, $I(\C;\G)$, is a
characteristic function of the data which does not depend on any
specific classification algorithm.

We are now ready to give a full formulation of the problem: {\em how
do we find classes of galaxies that maximize $I(\C;\Lm)$, under a
constraint on their complexity, $I(\C;\G)$?}  This constrained
information optimization problem was first presented in Tishby et al.
(1999) and their solution was termed the {\em
information bottleneck method}. In effect we pass the information that
$\G$ provides about $\Lm$ through a ``bottleneck'' formed by the
classes in $\C$. The classes $\C$ are forced to extract the relevant
information between $\G$ and $\Lm$.

Under this formulation, the {\em optimal} classification is given by
maximizing the functional
\begin{equation}
\label{varprin}
\L [p(\c|\g)] =  I(\C ; \Lm) - \beta^{-1} I(\C; \G) ,
\end{equation}
where $\bb$ is the Lagrange multiplier attached to the complexity
constraint.  For $\beta \rightarrow 0$ our classification is as
non-informative (and compact) as possible --- all galaxies are
assigned to a single class.  On the other hand, as $\beta \rightarrow
\infty$ the representation becomes arbitrarily detailed.  By varying
the single parameter $\beta$, one can explore the tradeoff between the
preserved meaningful information, $I(\C;\Lm)$, and the compression
level, $I(\C;G)$, at various resolutions.

Perhaps surprisingly, this general problem of extracting the relevant
information --- formulated in Eq. (\ref{varprin}) --- can be given an
{\em exact} formal solution.  In particular, the optimal assignment
that maximizes Eq. (\ref{varprin}) satisfies the equation
\begin{equation}
\label{r3}
p(\c|\g)
= {{p(\c)}\over Z(\g,\beta)}
\exp\left[ - \beta \sum_{\l} p(\l|\g) 
\log {{p(\l|\g)}\over{p(\l|\c)}}\right] ~,
\end{equation}
where $Z(g,\beta)$ is the common normalisation (partition) function.
\footnote{
We note that $\beta$ here is analogous to the inverse temperature in
the Boltzmann's distribution function.}  The value in the exponent can
be considered the relevant ``distortion function'' between the class
and the galaxy spectrum. It turns out to be the familiar cross-entropy
(also known as the `Kullback-Leibler divergence', e.g. Cover \& Thomas
1991), defined by
\begin{equation}
\label{KL}
D_{KL}\left[p(\l|\g) \Vert p(\l|\c)\right] = \sum_{\l} p(\l|\g) \log
{{p(\l|\g)}\over{p(\l|\c)}} ~.
\end{equation}
We emphasise that this effective distortion measure {\em emerges} 
here from first principles of information preserving
and was not imposed as an ad-hoc measure of spectral similarity.
Note that Eqs. (\ref{r1}, \ref{r2}, \ref{r3})
must be solved together in a {\em self-consistent} manner.

\subsection{Relations to conventional classification approaches}
\label{alg_simple}
We may gain some intuition into this method by contrasting
Eq. \ref{r3} with more standard clustering algorithms. Suppose we
start from Bayes' theorem, where the probability for a class $\c$ for
a given galaxy $\g$ is
\begin{equation}
p(\c|\g)  \propto 
p(\c) \; p(\g|\c)\;,  
\end{equation}
and $p(c)$ is the prior probability for class $c$.  As a simple ad-hoc
example, we can take the conditional probability $p(\g|\c)$ to be a
Gaussian distribution with variance $\sigma^2$
\begin{equation}
p(\g|\c)  = \frac{1}{\sqrt{2 \pi}\sigma}\; 
\exp (-\frac{1}{2 \sigma^2} D_{E}^{2})\;,  
\end{equation}
with
\begin{equation}
D_{E} = \sqrt{\sum_\l [p(\l|c) -p(\l|g)]^2}\;. 
\end{equation}
The Euclidean distance $D_{E}$ is analogous to the cross-entropy
$D_{KL}$ of Eq. \ref{KL}.  However, unlike our earlier formulation,
neither the distance $D_{E}$ nor the obtained classes have good
theoretical justifications.  The variance $\sigma^2$ (which may be due
cosmic scatter as well as noise) plays a somewhat analogous role to
the Lagrange multiplier $\beta$, but unlike $\beta$ it forces a fixed
`size' for all the clusters.  Hence $\sigma$ can be viewed as the
`resolution' or the effective `size' of the class in the
high-dimensional representation space.  We note that the Euclidean
distance is commonly used in supervised spectral classification using
`template matching' (e.g. Connolly et al. 1995; Benitez 1999), in
which galaxies are classified by matching the observed spectrum with a
template obtained either from a model or from an observed standard
galaxy. By comparing our method with an ``Euclidean algorithm'', we
find that our approach yields better class boundaries and preserves
more information for a given number of classes.
  
\subsection{The agglomerative information bottleneck algorithm} 
\label{alg}
The initial approach to the solution of the three self-consistent
Eqs. (\ref{r1}, \ref{r2}, \ref{r3}), applied already in Pereira, Tishby 
\& Lee (1993), was similar to the ``deterministic annealing'' method
(see e.g. Rose 1998).  This is a top-down hierarchical algorithm that
starts from a single class and undergoes a cascade of class splits
(through second order phase transitions) into a ``soft'' tree of
classes.  Here we use an alternative algorithm, first introduced in
Slonim \& Tishby (1999), based on a bottom-up {\em merging} process.
This algorithm generates ``hard'' classifications, i.e.  every galaxy
$\g \in \G$ is assigned to exactly one class $\c \in \C$. Therefore,
the membership probabilities $p(\c|\g)$ may only have values of $0$ or
$1$.  Thus, a specific class $\c$ is defined by the following
equations, which are actually the ``hard'' limit $\beta
\rightarrow \infty$ of the general self-consistent Eqs. (\ref{r1},
\ref{r2}, \ref{r3}) presented previously,
\beq{hard1}
\left\{
\begin{array}{l}
p(\c)  = \sum_{\g \in \c}{p(\g)} \\\\
p(\l|\c)  = \frac{1}{p(\c)} \sum_{\g \in \c} p(\l|\g)p(\g) \\\\
p(\c|\g) = \left\{ \begin{array}{ll} 1 & \mbox{if $\g \in \c$} \\ 0 &
      \mbox{otherwise} \end{array} \right. \\
\end{array}
\right.
\eeq
where for the second equation we used Bayes' theorem, $p(\g|\c) =
\frac{1}{p(\c)}p(\c|\g)p(\g)$.

The algorithm starts with the trivial solution, where $\C \equiv \G$
and every galaxy is in a class of its own. In every step two classes
are merged such that the mutual information $I(\C;\Lm)$ is maximally
preserved.  The merging process is formally described as follows.
Assume that we merge the two classes $\c_1,\c_2$ into a new class
$\c^{*}$. Then the equations characterizing the new class are
naturally defined by
\beq{merge}
\left\{
\begin{array}{l}
p(\c^{*})  = p(\c_1) + p(\c_2) \\\\
p(\l|\c^{*})  = \frac{p(\c_1)}{p(\c^{*})}p(\l|\c_1) + \frac{p(\c_2)}{p(\c^{*})}p(\l|\c_2) \\\\
p(\c^{*}|\g) = \left\{ \begin{array}{ll} 1 & \mbox{if $\g \in \{\c_1\} \bigcup \{\c_2\}$} 
        \\ 0 & \mbox{otherwise} \end{array} \right. \\
\end{array}
\right.
\eeq
The information loss with respect to $\Lm$ due to this merger is given
by
\begin{equation}
\label{delta}
\delta I(\c_1,\c_2) \equiv I(\C_{before};\Lm) - I(\C_{after};\Lm) \geq 0 ~,
\end{equation}
where $I(\C_{before};\Lm)$ and $I(\C_{after};\Lm)$ are the mutual
information between the galaxy classes and the wavelengths before and
after the merge, respectively.  Using
Eqs.(\ref{hard1},\ref{merge},\ref{delta}) it can be shown after some
algebra (Slonim \& Tishby, in preparation) that
\begin{equation}
\label{delta2}
\delta I(\c_1,\c_2) = 
(p(\c_1)+p(\c_2)) \cdot D_{JS} (\c_1,\c_2) ,
\end{equation}
where $D_{JS}$ is the {\em Jensen-Shannon} divergence (Lin 1991;
El-Yaniv, Fine \& Tishby 1997), defined by
\begin{equation}
\label{JS}
D_{JS} (\c_1,\c_2) = 
\sum_{i=1}^2 {p(\c_i) D_{KL}[p(\l|\c_i) \Vert 
\sum_{i=1}^2{p(\c_i) p(\l|\c_i)}]} ~ .
\end{equation}
An intuitive interpretation is that the ``merging cost'' (in
information terms) is equal to the ``distance'' $D_{JS}(\c_1,\c_2)$
between the classes before merging
\footnote{This distance has an interesting statistical interpretation as the 
distance to the most likely joint source of the two
classes. Alternately, it can be viewed as analogous to the physical
mixing entropy of two 
pure gases (see El-Yaniv et al. 1997; 
Bialek, Nemenman and Tishby
2000).}  multiplied by their ``weight'',
$p(\c_1)+p(\c_2)$.  The algorithm is now straightforward --- in each
step we perform ``the best possible merger'', i.e.  we merge the two
classes which minimize $\delta I(\c_i,\c_j)$. In this way, we maximize
$I(\C;\Lm)$ in every step (but note that this does not necessarily
guarantees a global maximum at the endpoint). In figure~\ref{fig:alg},
we give the pseudo-code for this procedure.  Note that this algorithm
naturally finds a classification for any desired number of classes
with no need to take into account the theoretical constraint via
$\beta$ (Eq. \ref{varprin}).  This is due to the fact that the
agglomerative procedure contains an inherent {\em algorithmic}
compression constraint, i.e. the merging process.  A more general
version of this algorithm which directly implements
Eq. (\ref{varprin}) is described elsewhere (Slonim \& Tishby, in
preparation).

For comparison with some conventional grouping algorithms, we also
implemented an algorithm which uses the Euclidean metric instead of
the Jensen-Shannon divergence used in Eq. (\ref{delta2}).  In this
case, in each step we merge the pair that minimizes $(p(\c_i)+p(\c_j))
\cdot \sqrt{\sum_{\l} (p(\l|\c_i)-p(\l|\c_j))^2}$, while ignoring the
statistical meaning of the distributions.  We refer to this procedure
as the {\em Euclidean} algorithm.

\begin{figure}
\centerline{\fbox{\begin{minipage}{20pc}
\underline{\large\bf Input}\\The empirical joint probability $p(g,\l) ~~ $\\ \\
\underline{\large\bf Output}\\A classification of $G$ into $N_{C}$ classes, for all  
                               $1 \leq N_{C} \leq N_{G}$\\ \\
\underline{\large\bf Initialization:} 
\begin{itemize}
\item  Construct $\C \equiv \G$
\item  $\forall i,j = 1...N_{G}, \; i < j$, calculate\\
       $~~~~~~~~ d_{i,j} = (p(\c_i)+p(\c_j)) \cdot D_{JS}(\c_i,\c_j) $\\
\end{itemize}
\underline{\large\bf Loop:}
\begin{itemize}
\item   For $N_{C} = N_{G}-1 ... 1$
\begin{itemize}
\item   Find the pair $(\c_i,\c_j)$ for which $d_{i,j}$ is minimized
\item   Merge $\{\c_{i}, \c_{j}\} \Rightarrow {\c_{*}}$ 
\item Update $\C = \{ \C - \{\c_{i},\c_{j}\} \} \bigcup \{ \c_{*} \}$
\item Update $d_{i,j}$ costs w.r.t. the new class $\c_{*}$
\end{itemize}
\item  {End For}
\end{itemize}
\end{minipage}}}
\caption{Pseudo-code of the agglomerative information bottleneck algorithm.}
\label{fig:alg}
\end{figure}

\section{Spectral Ensembles}
\label{sec:data}
\subsection{Observed Spectra from the 2dF Survey}
\label{sec:data:2df}
The 2dF Galaxy Redshift Survey (2dFGRS; Colless 1998, Folkes et al. 1999) is
a major new redshift survey utilising the 2dF multi-fibre spectrograph
on the Anglo-Australian Telescope (AAT). The observational goal of the
survey is to obtain high quality spectra and redshifts for 250,000
galaxies to an extinction-corrected limit of $b_J$=19.45. The survey
will eventually cover approximately 2000 sq deg, made up of two
continuous declination strips plus 100 random $2^\circ$-diameter
fields.  Over 100,000 galaxy spectra have been obtained as of May 
2000.  The spectral scale is 4.3\AA per pixel and the FWHM resolution
is about 2 pixels.  Galaxies at the survey limit of $b_J$=19.45 have a
median S/N of $\sim 14$, which is more than adequate for measuring
redshifts and permits reliable spectral types to be determined.

Here we use a subset of 2dF galaxy spectra, previously used in the
analysis of Folkes et al. (1999).  We emphasize that the spectra are
left in terms of {\em photon counts} (as opposed to energy flux). The
spectra were de-redshifted to their rest frame and re-sampled to a
uniform spectral scale with 4\AA\ bins. Since the galaxies cover a
range in redshift, the rest-frame spectra cover different wavelength
ranges. To overcome this problem, only objects with redshifts in the
range $0.01 \le z \le 0.2$ were included in the analysis.  All the
objects meeting this criterion then have rest-frame spectra covering
the range 3700\AA\ to 6650\AA\ (the lower limit was chosen to exclude
the bluest end of the spectrum where the response function is poor).
Limiting the analysis to this common wavelength range means that all
the major optical spectral features between O$_{\rm II}$ (3727\AA) and
H$\alpha$ (6563\AA) are included. In order to make the spectral
classifications as robust as possible, objects with low S/N were
eliminated by imposing a minimum mean flux of 50 counts per bin. The
spectra were then normalised so that the mean flux over the whole
spectral range was unity. The final sample contains 5869 galaxies,
each described by 738 spectral bins. Throughout this paper, we refer
to this ensemble as the ``2dF catalogue''.

We corrected each spectrum by dividing it by a global system response
function (Folkes et al. 1999).  However, it is known that due to
various problems related to the telescope optics, the seeing, the
fibre aperture etc.  the above correction is not perfect. In fact,
each spectrum should be corrected by an individual response function.
Unfortunately this incomplete correction mainly affects the continuum
of the spectrum, i.e. the galaxy `colour'.  As we shall see below, the
IB analysis on the mock data (which obviously is free of the above
problems) shows that the colour is a significant indicator of the
underlying astrophysics.  This highlights the need to correct properly
each individual spectrum (work in progress).

\subsection{Model Spectra from Semi-Analytic Hierarchical Merger Models}
\label{sec:data:models}
Our goal is to produce an ensemble of synthetic spectra with a
representative admixture of different types of galaxies with realistic
star formation histories. One way to accomplish this is to use
semi-analytic modelling techniques (cf. Kauffmann, White, \&
Guiderdoni 1993; Cole et al. 1994; Somerville \& Primack 1999 (SP) and
references therein). Semi-analytic models have the advantage of
being computationally efficient, while being set within the
fashionable hierarchical framework of the Cold Dark Matter (CDM)
scenario of structure formation. In addition to model spectra, this
approach provides many physical properties of the galaxies, such as
the mean stellar age and metallicity, size, mass, bulge-to-disk ratio,
etc. This allows us to determine how effectively a given method can
extract this type of information from the spectra, which are
determined in a self-consistent way. We have used the code
developed by Somerville (1997), which has been shown to produce good
agreement with many properties of local (SP) 
and high-redshift (Somerville, Primack \& Faber 2000; SPF)
galaxies. Below  we briefly summarize the models.

The formation and merging of dark matter halos as a function of time
is represented by a ``merger tree'', which we construct using the
method of Somerville \& Kolatt (1999). The number density of halos of
various masses is determined by an improved version of the
Press-Schechter model (Sheth \& Tormen 1999), which mostly cures the
usual discrepancy with N-body simulations. The cooling of gas,
formation of stars, and reheating and ejection of gas by supernovae
within these halos is modelled by simple recipes. Chemical evolution
is traced assuming a constant yield of metals per unit mass of new
stars formed. Metals are cycled through the cold and hot gas phase by
cooling and feedback, and the stellar metallicity of each generation
of stars is determined by the metal content of the cold gas at the
time of its formation. All cold gas is assumed to initially cool into,
and form stars within, a rotationally supported disk; major mergers
between galaxies destroy the disks and create spheroids. New disks may
then be formed by subsequent cooling and star formation, producing
galaxies with a range of bulge-to-disk ratios. Galaxy mergers also
produce bursts of star formation, according to the prescription
described in SPF. Thus the star formation history of a single galaxy
is typically quite complex and is a direct consequence of its
environment and gas accretion and merger history.

These star formation histories are convolved with stellar population
models to calculate magnitudes and colors and produce model
spectra. We have used the multi-metallicity GISSEL models (Bruzual \&
Charlot, in preparation) with a Salpeter IMF to calculate the stellar part of
the spectra. Emission lines from ionized $H_{\rm II}$ regions have
been added using the empirical library included in the PEGASE models
(Fioc \& Rocca-Volmerange 1997). This library provides the total
luminosity of all of the major emission lines as a function of the age
of the stellar population. Any dependence of the line strengths on
metallicity, ionization state, geometry, etc. is neglected. We adjusted
the resolution of the mock spectra to be comparable to the 2dF spectra. The
width of the lines is then determined by the resolution of the
grating, and we have modelled them as Gaussians with a width of 4
\AA. 

Dust extinction is included using an approach similar to that of
Guiderdoni \& Rocca-Volmerange (1987). Here, the mass of dust is
assumed to be a function of the gas fraction times the metallicity of
the cold gas. We then use a standard Galactic extinction curve and a
``slab'' model to compute the extinction as a function of wavelength
and inclination (see Somerville 1997 or Somerville et al. 2000b for
details). The extinction correction is applied indiscriminantly to the
stellar and line emission. This is probably unphysical as it is likely
that the star-forming regions that produce the emission lines are more
heavily extinguished than the underlying old stellar population, but
at our current level of modelling, we ignore this effect.

As described in SP, we set the free parameters of the models by
reference to a subset of local galaxy data; in particular, we require
a typical $L_{*}$ galaxy to obey the observed I-band Tully-Fisher
relation and to have a gas fraction of $0.1$ to $0.2$, consistent with
observed gas contents of local spiral galaxies. If we assume that
mergers with mass ratios greater than $\sim$ 1:3 form spheroids, we
find that the models produce the correct morphological mix of spirals,
S0s and ellipticals at the present day (we use the mapping between
bulge-to-disk ratio and morphological type from Simien \& de
Vaucouleurs 1986).  This critical value for spheroid formation is what
is predicted by N-body simulations of disk collisions (cf. Barnes \&
Hernquist 1992).

In SP, we found that a cosmology with $\Omega_0=\Omega_{\Lambda}=0.5$
and $H_0 = 60$ km/s/Mpc produced very good agreement with the 2dF
$b_J$-band luminosity function (for all types combined), as well as
the observed K-band luminosity function, Tully-Fisher relation,
metallicity-luminosity relation, and colours of local galaxies. We use
the same fiducial model here, with a few minor modifications: we
incorporate self-consistently the modelled metallicity of the hot gas
in the cooling function, and use the multi-metallicity SED's (instead
of solar metallicity) with a Salpeter (instead of Scalo) IMF. Another
minor detail is that ejected material is eventually returned to the
halos as described in the updated models of SPF. We find that these
minor modifications do not significantly change our previous results
for local galaxies.

We construct a ``mock 2dF catalogue'' of $2611$ model galaxies with the same
magnitude limit, wavelength coverage and spectral resolution, and
redshift range as the 2dF survey (described above).
The synthetic spectra are expressed in terms of photon
counts and the total number of counts in each spectrum is normalized
to unity, as in the prepared observed spectra.
We shall present elsewhere comparison of the mock catalogues with preliminary data from
the 2dF survey

To study the effect of noise on the classification, we added Poisson
noise to the simulated spectra.  This was done using an empirical
relation between the mean photon counts ${\bar N_{ph}}$ per observed
2dF spectrum and the corresponding APM $b_J$ magnitude, $ {\bar
N_{ph}} = 9638.0 - 757.8 \;b_J + 13.9 \;b_J^2$ (D. Madgwick, private
communication).  Each simulated galaxy was assigned a mean number of
counts based on its $b_J$ magnitude as given by the models, and
Poisson deviants per spectral bin were drawn at random.
\footnote{
We note that an alternative approach could be to filter out the noise of the
observed spectra, e.g. by using PCA (Folkes et al. 1996).}  
We have ignored the effects of the response function of the fibres, aperture
effects, and systematic errors related to the placement of fibres in
the holding plate (see above).

\begin{figure}
\centerline{
\psfig{file=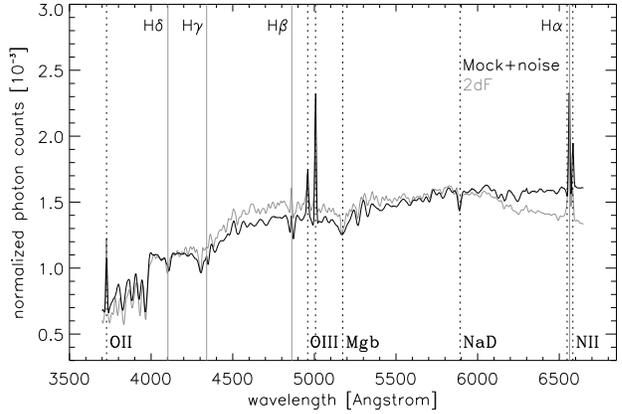,width=9cm,height=6cm}}
\caption{Photon counts (sum normalised to unity) 
as a function of wavelength, averaged over the
entire catalogues of 2dF and mock+noise spectra. Note the familiar
spectral features such as the Balmer break at 4000 \AA, the Balmer
series (marked with vertical grey lines), and metal lines such as O,
Mg, Na, and N (marked by dotted lines).
\label{fig:meanspec}}
\end{figure}
Figure~\ref{fig:meanspec} shows the mean spectrum for the 2dF and
mock+noise catalogues, obtained by simply averaging the photon counts
in each wavelength bin for all the galaxies in the ensemble. The mean
spectra for the observed and mock catalogues are seen to be
similar. The magnitude limit that we have chosen is such that our
ensembles are dominated by fairly bright, moderately star-forming
spiral galaxies, and the mean spectra show familiar features such as
the 4000 \AA\ break, the Balmer series, and metal lines such as
O$_{\rm II}$ and O$_{\rm III}$. One can see that the 2dF spectrum
appears to bend downwards relative to the models towards both ends of
the wavelength range. This may be due to an inaccurate correction for
the response function, as discussed above.

\section{Results}
\begin{figure*}
\centerline{
\psfig{file=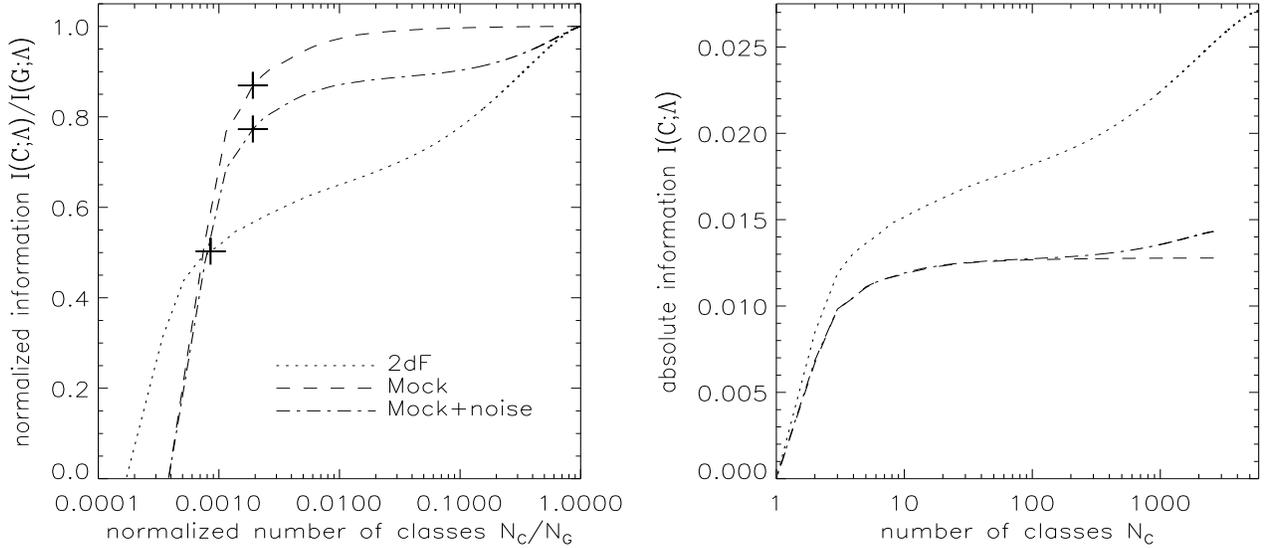,width=18cm,height=8cm}}
\caption
{Left Panel: The fractional information measure
($I(\C;\Lm)/I(\G;\Lm)$) vs. the fractional number of classes
($N_{C}/N_{G}$). The crosses mark the points corresponding to five
classes, used in the remainder of this paper.  Right Panel: The
`absolute' information content $I(\C;\Lm)$ as a function of the number
of classes $N_C$.  }
\label{fig:info}
\end{figure*}

We now apply the IB algorithm to both the 2dF and the mock data.
Recall that our algorithm begins with one class per galaxy, and groups
galaxies so as to minimize the loss of information at each
stage. Figure~\ref{fig:info} shows how the information content of the
ensemble of galaxy spectra decreases as the galaxies are grouped
together and the number of classes decreases. In the left panel, we
show the `normalized' information content $I(\C;\Lm)/I(\G;\Lm)$ as a
function of the reduced complexity $N_{C}/N_{G}$, where $N_{G}$ is the
number of galaxies in the ensemble and $N_{C}$ is the number of
classes.  One may think of this as the fraction of information about
the original ensemble that would be preserved if we threw away all the
individual galaxies and kept only the representative spectra of the
classes. Remarkably, we find that if we keep about five classes, about
$85$ and $75$ percent of the information is preserved for the mock and
mock+noise simulations, respectively. This indicates that the
wavelength bins in the model galaxy spectra are highly correlated.
This may not seem very surprising, since many of the spectral features
arise from the same stellar physics. For example, emission lines will
be stronger in a galaxy with significant recent star formation, and
the whole sequence of metal absorption lines will be deeper for a
galaxy with an old stellar population or a high metallicity. However,
if we keep in mind that each of our model galaxies has a very complex
star formation history and is therefore comprised of stars with a
distribution of ages and metalicities, and is affected differently by
dust extinction, this result may seem somewhat more surprising.

In contrast with the mock samples, for the 2dF catalogue, only about
$50$ percent of the information is preserved by five classes. Thus the
wavelength bins in the real spectra are also highly correlated, but
not to the same degree as the model spectra. This is unlikely to be
solely due to the effect of noise.  Adding noise to the mock
catalogues does change the curve in Figure~\ref{fig:info} (left
panel), bringing it closer to the 2dF curve, but the effect is not
large enough to explain the whole discrepancy, and moreover the shapes
of the mock+noise and 2dF information curves are still different.  We
also see that the `absolute information' for 2dF (right panel of
Figure~\ref{fig:info}) is much higher than for the mock samples. This
discrepancy may be partially due to the influence on the real spectra
of more complicated physics than what is included in our simple
models. It could also be due to systematic observational errors (see
Section~\ref{sec:data:2df}). We are in the process of attempting to
model these systematic errors in detail to better understand this
result.

The information curves are not sensitive to the number of galaxies
used in the analysis. 
When we take a random subset of 2611 galaxies out of the 2dF sample
(to make it identical to the number of galaxies in the mock sample),
the differences in the information curve w.r.t the 
full 2dF set are minor.
We also experimented with the `Euclidean' algorithm (see
Section~\ref{alg_simple}) for both the mock and 2dF data. We find that
for the mock data, the Euclidean algorithm produces nearly
indistinguishable results from our fiducial algorithm, however, for
the 2dF data, the difference between the two is more significant. The full IB
algorithm preserves more information for any given number of classes
and is therefore superior. This suggests that the Euclidean-type
algorithms may be sufficient for certain problems, but inadequate for
more complex data.

\subsection{The IB Classes}
\begin{figure*}
\centerline{
\epsfig{file=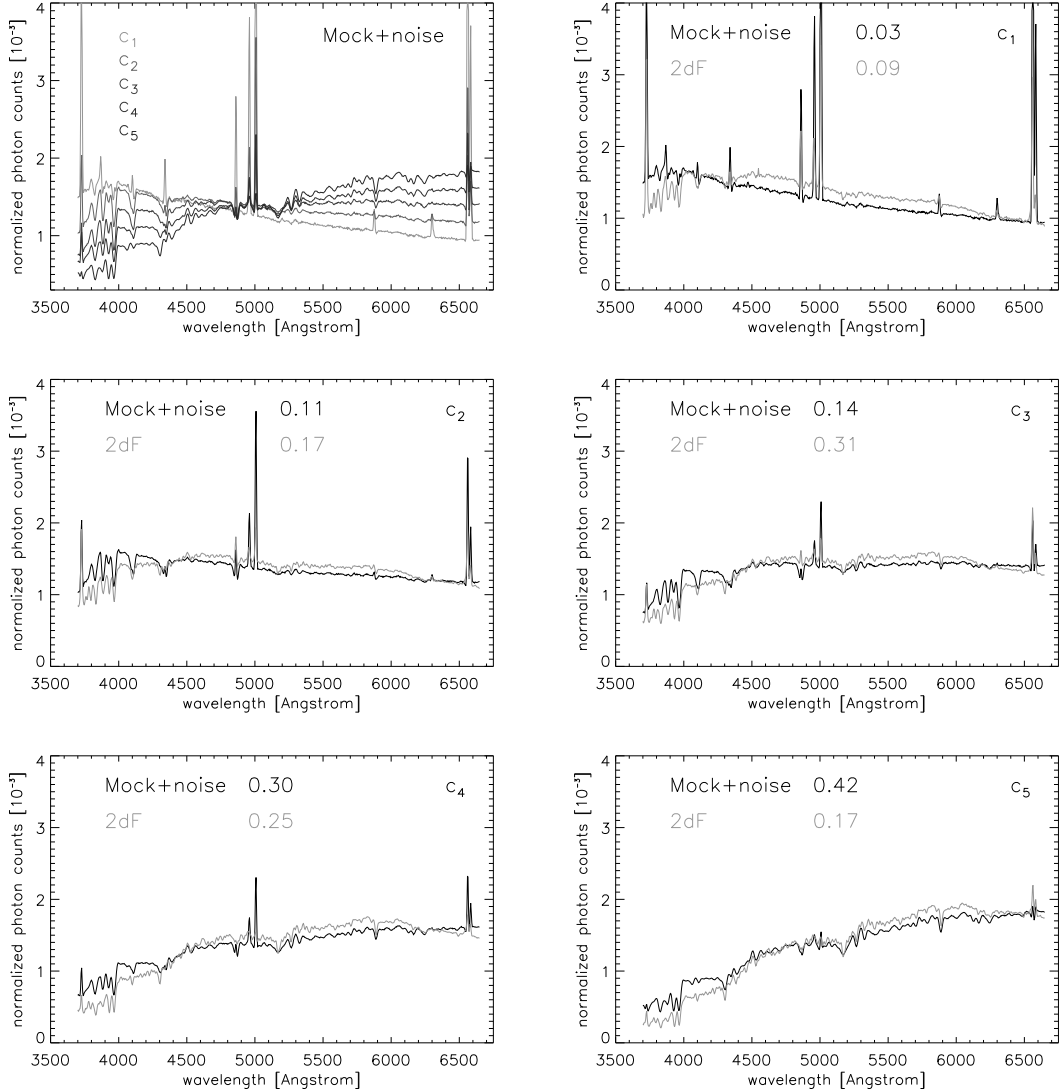,width=15truecm,height=15truecm}}
\caption{Mean spectra of the five IB classes for the 2dF and mock+noise
catalogues. The fraction $p(c_j)$ of galaxies that are members of each
class $c_j$ is indicated. The matching between the classes obtained
for 2dF and the mock catalogue was determined by minimizing the
average $D_{JS}$ `distance' between the pairs.
\label{fig:spectra5class}}
\end{figure*}

For the remainder of this paper, we present the results obtained for
five classes.  
\footnote {We note that 
galaxy images can be reliably classified 
by morphology 
into no more than 7 or so classes
(e.g. Lahav et al. 1995; Naim et al. 1995a) }
Figure~\ref{fig:spectra5class} shows the representative
spectra for these five classes for both the 2dF and mock+noise
catalogues. The corresponding five spectra for the noise-free mock
data were very similar to the mock+noise spectra shown. We `matched'
each of the classes obtained for the 2dF data with one from the
mock+noise data by minimizing the average $D_{JS}$ `distance' between
the pairs. The classes are then ordered by their mean $B-V$
colour. Note that the five classes produced by the algorithm appear
similar for both catalogues --- there was certainly no guarantee that
this would be the case.

It is also interesting to examine the relative fractions of galaxies
in each class, $p(c)$, for the observed and mock catalogues. These
values are given on the appropriate panels of
Figure~\ref{fig:spectra5class}. There is a partial agreement between
the weights of the matched classes for the 2dF and mock+noise
catalogues (although particularly for $\c_3$ and $\c_5$, the agreement
is not very good).  We might hope that this could provide a way to
improve the physics included in the models by constraining the
relative composition of different types of galaxies in more detail
than was previously possible. 
However, we find that adding noise to the mock catalogues causes some
galaxies to ``jump'' to different classes, thus changing the relative
fractions. When we add noise at the level of the 2dF data as described
previously, $\sim 23$ percent of the galaxies are assigned to a different
class than in the noise-free case. Of these, the great majority ($\sim 22$
percent) are assigned to adjacent classes (i.e. $\Delta c = \pm 1$).
We note these results are
for the `hard' version of the  algorithm.
We expect the `soft' version of the algorithm to be less sensitive
to noise. 
 
We are also concerned that some of the
noticeable discrepancy in the shape of the mean spectra of the 2dF
classes and those of the synthetic galaxies (also noted in the
comparison of the mean spectra, Figure~\ref{fig:meanspec}) may be due
to systematic observational effects such 
as inaccurate modelling of the response
function. In the future we hope to be able to model and correct for
these effects.

More generally, we can see that the algorithm is sensitive to the
overall slope (or colour) of the spectrum, and also to the strength of
the emission lines. The classes clearly preserve the familiar physical
correlation of colour and emission line strength; the five classes
form a sequence from $\c_1$, which has a blue continuum with strong
emission lines, to $\c_5$, with red continuum with no emission lines
\footnote{Recall that the order of the classes as produced by the
algorithm is arbitrary, and we have placed them in this sequence by
hand, but the correlation of colour and emission line strength within
the classes is produced by the algorithm with no help from
us}. Already, we may form the impression that the algorithm has
classified the galaxies in a way that is reminiscent of conventional
spectral classes. It may be interesting to compare the mean spectrum
of $\c_1$ with the spectrum of the Sm/Irr galaxy NGC449, and $\c_5$
with the Sa galaxy NGC775 from Figure 2a of Kennicutt (1992).
Apparently, the $\c_1$ class corresponds to late type galaxies
(Sm/Irr) and $\c_5$ to early types (Sa-E).

\subsection{Correlation of the IB Classes with Physical Properties}
\begin{figure*}
\centerline{
\psfig{file=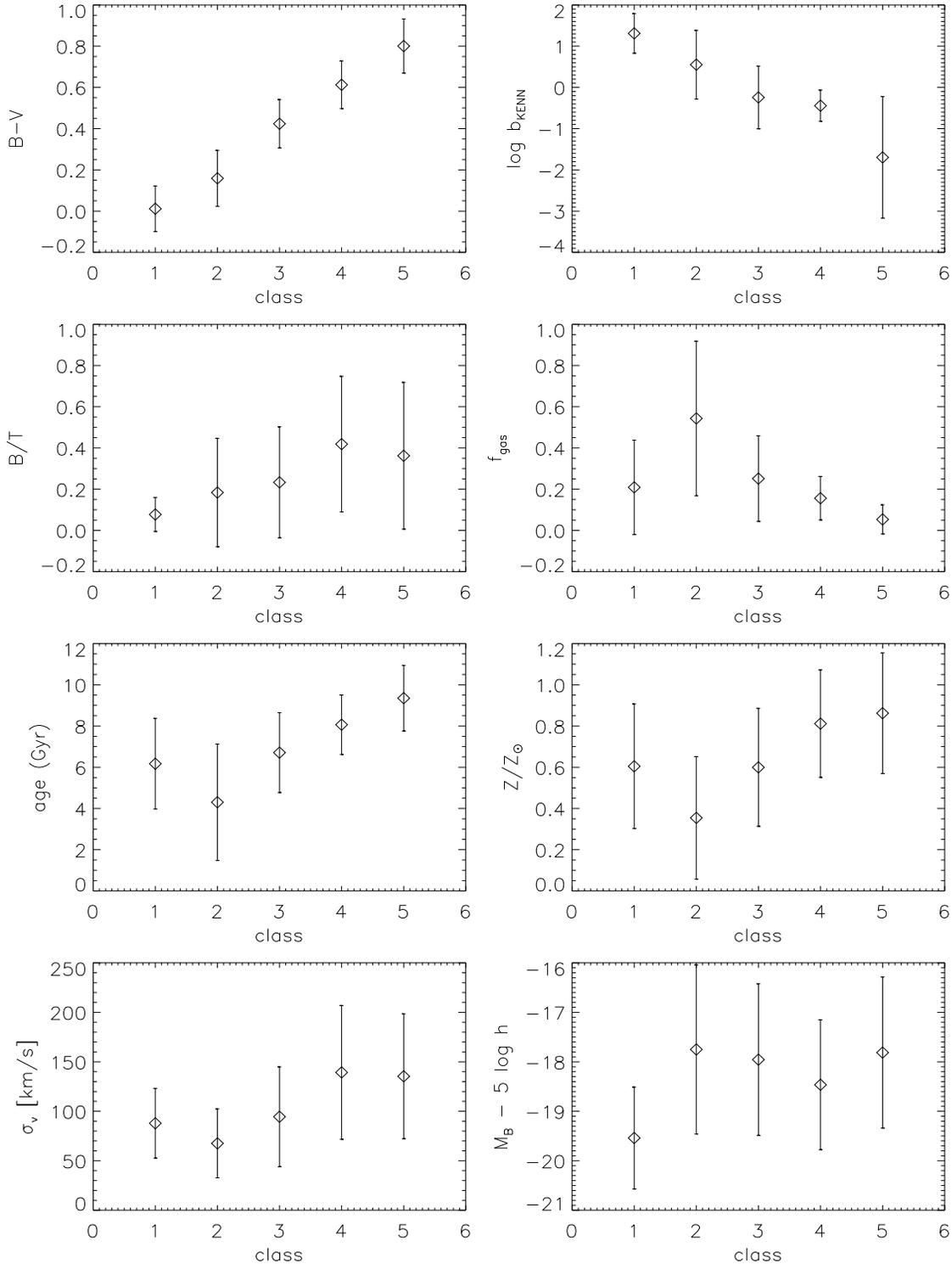,width=15cm,height=20cm}}
\caption{The trends of physical parameters (see text) with the five IB classes.
The error bars are 1-sigma. Strong trends of colour and
present-to-past-averaged star formation rate ($b_{\rm KENN}$) are
seen. Weaker trends of bulge-to-total ratio ($B/T$), mean stellar age,
gas fraction ($f_{\rm gas}$), stellar metallicity ($Z$), and internal
velocity dispersion ($\sigma_v$) are also apparent.
\label{fig:trends}}
\end{figure*}

In order to gain a better understanding of the IB classes, we now use
the noise-free mock catalogue and investigate the physical properties
of the galaxies in each class as given by the same models that we use
to produce the spectra. The parameters that we chose to investigate
are: $B-V$ colour, the ratio of the present to past-averaged star
formation rate (Kennicutt b-parameter, $b_{KENN} \equiv {SFR}/\langle
SFR\rangle$; Kennicutt 1983), the ratio of the mass of the bulge to
the total stellar mass of the galaxy ($B/T$), the fraction of the
total baryonic mass in cold gas ($f_{\rm gas} \equiv m_{\rm cold}/(
m_{\rm cold}+m_{\rm star})$), the mean mass-weighted stellar age and
metallicity, the internal velocity dispersion $\sigma_v$ or circular
velocity $V_c$, and the B-band absolute magnitude $M_B$.

Figure~\ref{fig:trends} shows the trends of these physical parameters
with class number. The strongest dependence is of $B-V$ colour and
present-to-past-averaged star formation rate $b_{\rm KENN}$. Class $\c_1$
contains blue galaxies that are forming stars at rates that are one to
two orders of magnitude higher than the average over their past
history. As one moves towards $\c_5$, galaxies are redder and formed a
larger fraction of their stars in the past. This is consistent with
the observed strong correlation of the Kennicutt b-parameter with
$B-V$ colour and morphological type in nearby galaxies (Kennicutt et
al. 1994). For reference, note that the mean colour for Sm/Irr
galaxies in the local Universe is $B-V=0.42$, for Sbc-Sc $B-V=0.55$,
for Sab-Sb $B-V=0.64$, for S0a-Sa $B-V=0.78$, and for E-S0 $B-V=0.90$
(Roberts \& Haynes 1994). From the colours alone, we might guess that
galaxies in $\c_1$ and $\c_2$ are starburst galaxies, galaxies in $\c_2$ are
Sm-Irr, $\c_4$ corresponds to Sb/Sbc and $\c_5$ to Sa/S0/E. This bolsters our
initial impression based on the visual appearance of the
spectra. Weaker trends are visible in other properties: the sequence
$\c_1$-$\c_5$ shows an increasing B/T ratio, as expected, but with a large
scatter within each class. There is quite a large scatter in the
observed correspondence between morphological T-type and
bulge-to-total ratio, but values of $B/T \sim 0.4-0.5$ are typical of
very early-type spirals (Sa) or lenticular galaxies (cf. Simien \& de
Vaucouleurs 1986). Weak trends are also visible in the mean stellar
age and metallicity and the internal velocity dispersion: as we move
from $\c_1$--$\c_5$, galaxies tend to be older, more metal rich, and more
massive. This is in accord with the observed trends of these
quantities with morphological type (cf. Roberts \& Haynes 1994) and
the trends we have noted previously. 

We find no trend of dust extinction with the IB classes, probably because we
have assumed that dust extinction attenuates the emission lines and the
continuum by the same factor. Therefore the dust only changes the
overall colour of a galaxy, but does not affect the ratio of emission
or absorption lines to continuum.

\begin{figure}
\centerline{\epsfig{file=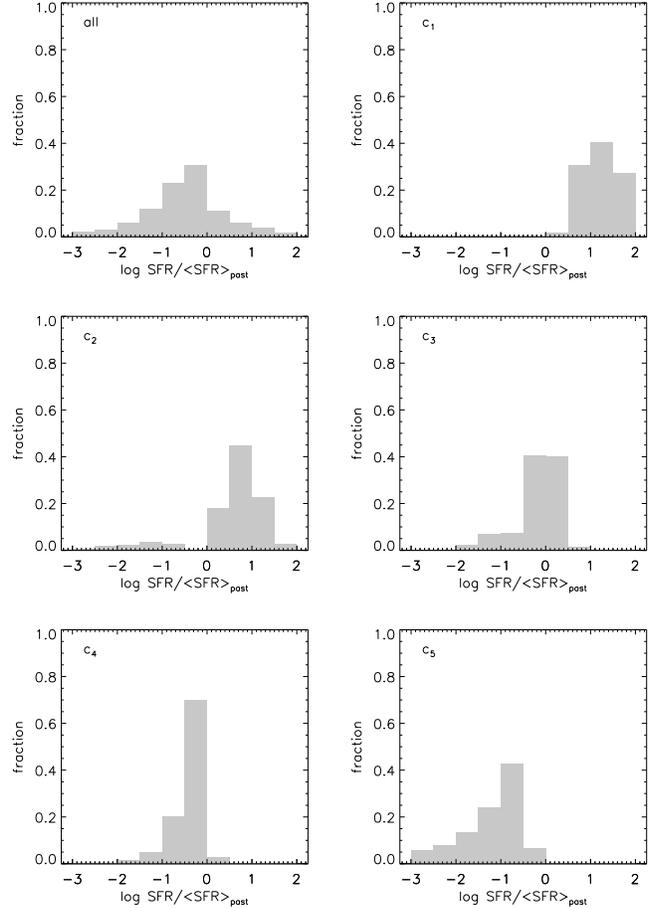,width=9truecm,height=12.5truecm}}
\caption{Distribution of the present over past-averaged star formation rate 
(Kennicutt's $b_{KENN}$ parameter) for the model galaxies. The top
left panel shows the distribution for all galaxies, and the remaining
panels show separately the distributions for galaxies in each of the
five IB classes.
\label{fig:sfrhist}}
\end{figure}

\begin{figure}
\centerline{\epsfig{file=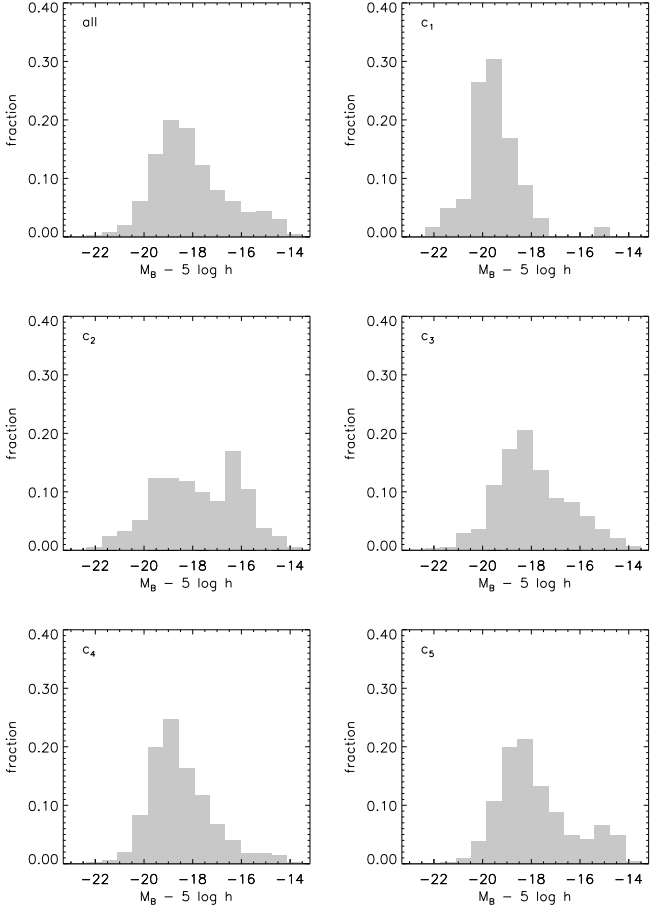,width=9truecm,height=12.5truecm}}
\caption{Distribution of absolute B-band magnitudes of the model galaxies 
(as in figure~\protect\ref{fig:sfrhist}). 
\label{fig:bmaghist}}
\end{figure}

\begin{figure}
\centerline{\epsfig{file=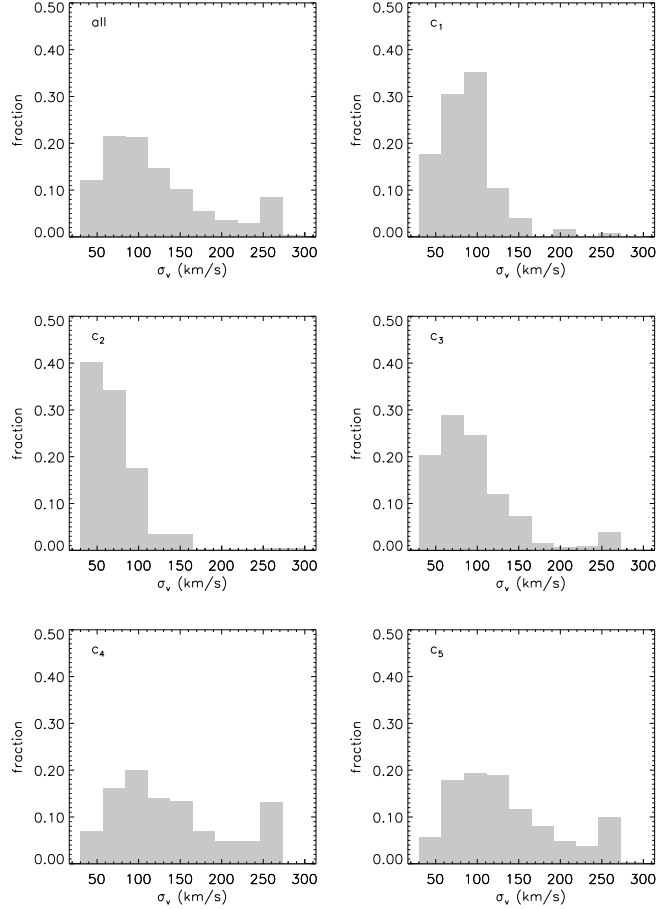,width=9truecm,height=12.5truecm}}
\caption{Distribution of internal velocity dispersion, $\sigma_v$, for the model galaxies
(as in figure~\protect\ref{fig:sfrhist}).
\label{fig:vhist}}
\end{figure}

\begin{figure}
\centerline{\epsfig{file=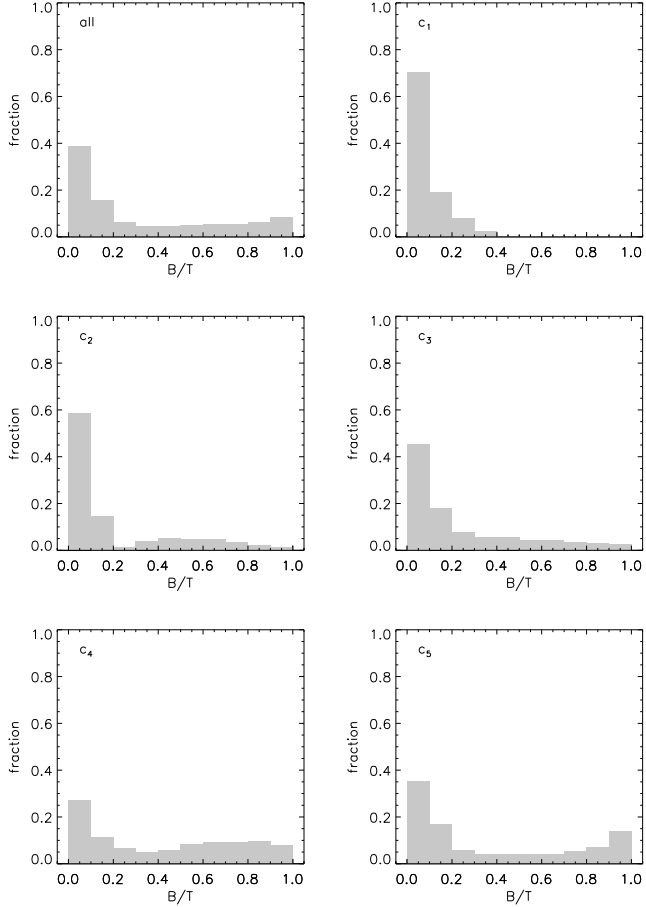,width=9truecm,height=12.5truecm}}
\caption{Distribution of the ratio of bulge to total stellar mass, for
the the model galaxies (as in
figure~\protect\ref{fig:sfrhist}). Galaxies with $B/T>0.4$--0.5 may
be associated with early type (spheroidal) galaxies.
\label{fig:btthist}}
\end{figure}

\begin{figure}
\centerline{
\epsfig{file=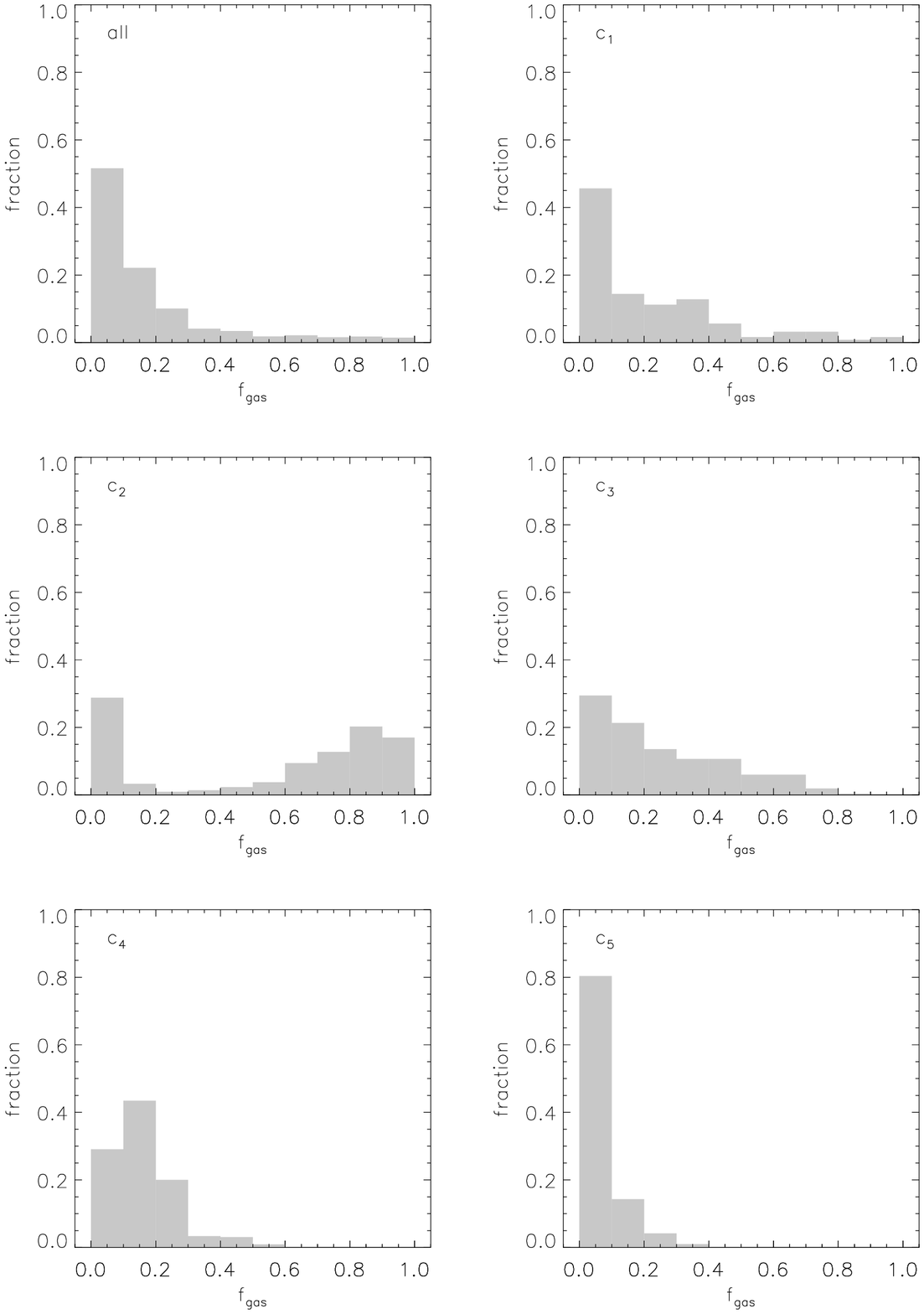,width=9truecm,height=12.5truecm}}
\caption{Distribution of the gas fraction for the model galaxies (as in
figure~\protect\ref{fig:sfrhist})
\label{fig:fgashist}}
\end{figure}

We investigate the composition of the classes in more detail by
examining the distributions of some of the physical parameters for the
different classes. In each of
figures~\ref{fig:sfrhist}-\ref{fig:fgashist}, in the upper left-most
panel we show the distribution of the relevant physical parameter for
the whole ensemble of model (noise-free) galaxies. In the other five
panels we show the distributions for the galaxies in each class
separately. Figure~\ref{fig:sfrhist} shows the distribution of
Kennicutt's $b_{KENN}$ parameter (ratio of present-to-past-averaged
SFR). This parameter shows a particularly strong correlation with the
IB class assignment, although the distributions show significant
overlap. Figure~\ref{fig:bmaghist} shows the distribution of B-band
absolute magnitudes. Class $\c_1$ contains an excess of bright
galaxies, whereas $\c_2$ and to a lesser extent $\c_5$ contain
excesses of faint galaxies compared to the overall distribution.
Figure~\ref{fig:vhist} shows the distribution of the internal velocity
dispersion $\sigma_v$ of the galaxies, a measure of the dynamical mass
of the galaxy. Here we note the curious fact that the distribution is
skewed towards small $\sigma_v$ galaxies in $\c_1$-$\c_3$, and towards
large $\sigma_v$ galaxies in $\c_4$-$\c_5$. This suggests that
galaxies in $\c_1$, which are preferentially {\em bright} and with
{\em small mass} are starburst galaxies. We return to this point
later.

In Figure~\ref{fig:btthist}, we see that $\c_1$ is entirely composed of
disk-dominated galaxies ($B/T \ltsim 0.4$). The classes become
progressively more skewed towards larger $B/T$, bulge-dominated
galaxies as we move towards $\c_5$. However, despite the fact that we
noted that the spectral appearance, colors and star formation rates of
galaxies in $\c_4$-$\c_5$ are typical of observed galaxies with bulges (Sb-E),
in the models these classes contain a significant fraction of nearly
pure disks ($B/T \ltsim 0.2$). This may be an indication that the
connection between star formation and morphology is not being modelled
properly.

The distribution of gas fraction is shown in
Figure~\ref{fig:fgashist}.  In $\c_1$, there are quite a lot of galaxies
with fairly low gas fractions. This is somewhat in conflict with our
previous association of this class with late-type galaxies, but
perhaps not if most of these galaxies have experienced a very recent
starburst which would have tended to consume the gas supply very
quickly. The class $\c_2$ is highly skewed towards high gas fractions, but with some
objects with low gas fractions --- perhaps this class is composed of a
combination of post-starburst galaxies and quiescent, gas rich
galaxies. Classes $\c_4$-$\c_5$ are composed solely of rather gas-poor
galaxies, confirming that these galaxies have probably been evolving
passively, with little new star formation.

\begin{figure}
\centerline{\epsfig{file=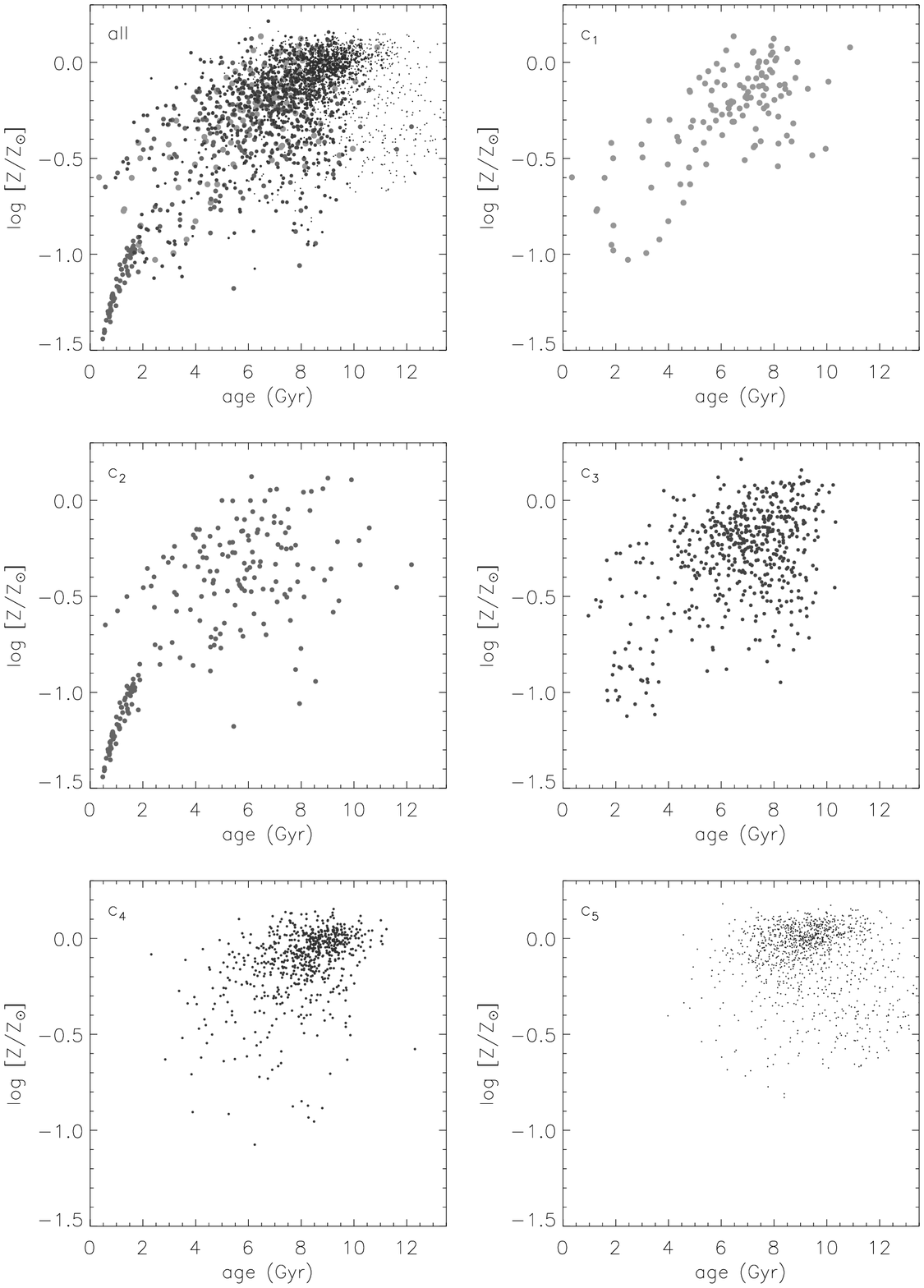,width=9truecm,height=12.5truecm}}
\caption{Stellar metallicity vs. mean stellar age for the model
galaxies. The top left panel shows all galaxies, and the remaining
panels show separately the galaxies in each of the five IB classes.
\label{fig:agemet}}
\end{figure}

It is also useful to see where the classes are located in the
two-dimensional space of pairs of the physical parameters. In the
following figures, we once again show the entire ensemble in the upper
left panel, and the breakdown by class in the other five panels.
Figure~\ref{fig:agemet} shows the mean stellar metallicity of galaxies
as a function of their mean stellar age.  We see that there is a weak
trend, with a large scatter, between these two quantities in our
models. This trend is stronger for classes $\c_1$-$\c_3$, and becomes
mostly washed out for $\c_4$-$\c_5$. Galaxies in $\c_1$-$\c_2$ also
tend to have higher metallicities for their age than galaxies in
$\c_4$-$\c_5$. As applied here, the algorithm suffers from the
familiar age-metallicity degeneracy. However, an alternative way of
applying the algorithm (by asking it to preserve the maximum
information {\em about a particular physical parameter}, e.g. age or
metallicity) may prove to be effective for this problem. We intend to
pursue this in future work.

\begin{figure}
\centerline{\epsfig{file=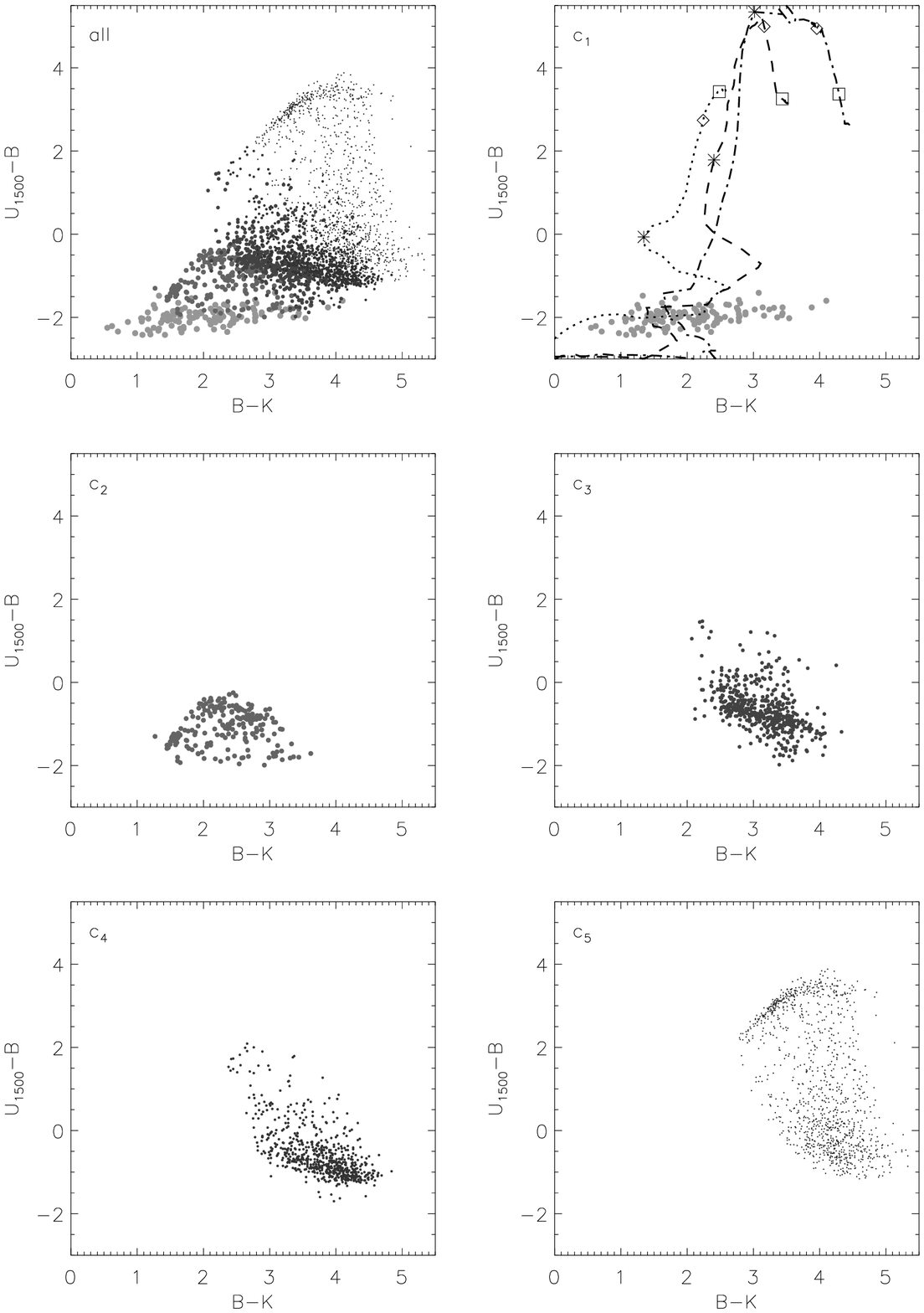,width=9truecm,height=12.5truecm}}
\caption{$U_{1500}-B$ colour vs. B-K colour diagram for the model galaxies.
The lines in the top right panel show the tracks of single-age and
metallicity instantaneous bursts from the Bruzual \& Charlot
models. Symbols mark single-burst ages of 1 Gyr (star), 5 Gyr
(diamond), and 10 Gyr (square). Three different metallicities are
shown: 0.005 $Z_{\odot}$ (dotted), 0.2 $Z_{\odot}$ (dashed), and solar
(dot-dashed).
\label{fig:colcol}}
\end{figure}

We also examine how the classes inhabit a classical colour-colour
diagram. Figure~\ref{fig:colcol} shows the far-UV/optical (1500 \AA
-B) and optical/near-IR (B-K) colours. Recall that the spectra that we
provided to the IB algorithm did not contain any information about the
far-UV (1500 \AA) or near-IR (K-band) magnitudes. Galaxies in $\c_1$
have a narrow range of very blue UV-B colours but a broad range of B-K
colours. The shape and orientation of this clump changes as we
move from $\c_1$-$\c_5$. The lines on the top right panel show the
tracks for instantaneous bursts of fixed age and metallicity from the
Bruzual \& Charlot models (see figure caption). This illustrates the
complex manner in which both age and metallicity determine the
location of galaxies in this diagram.

\begin{figure}
\centerline{\epsfig{file=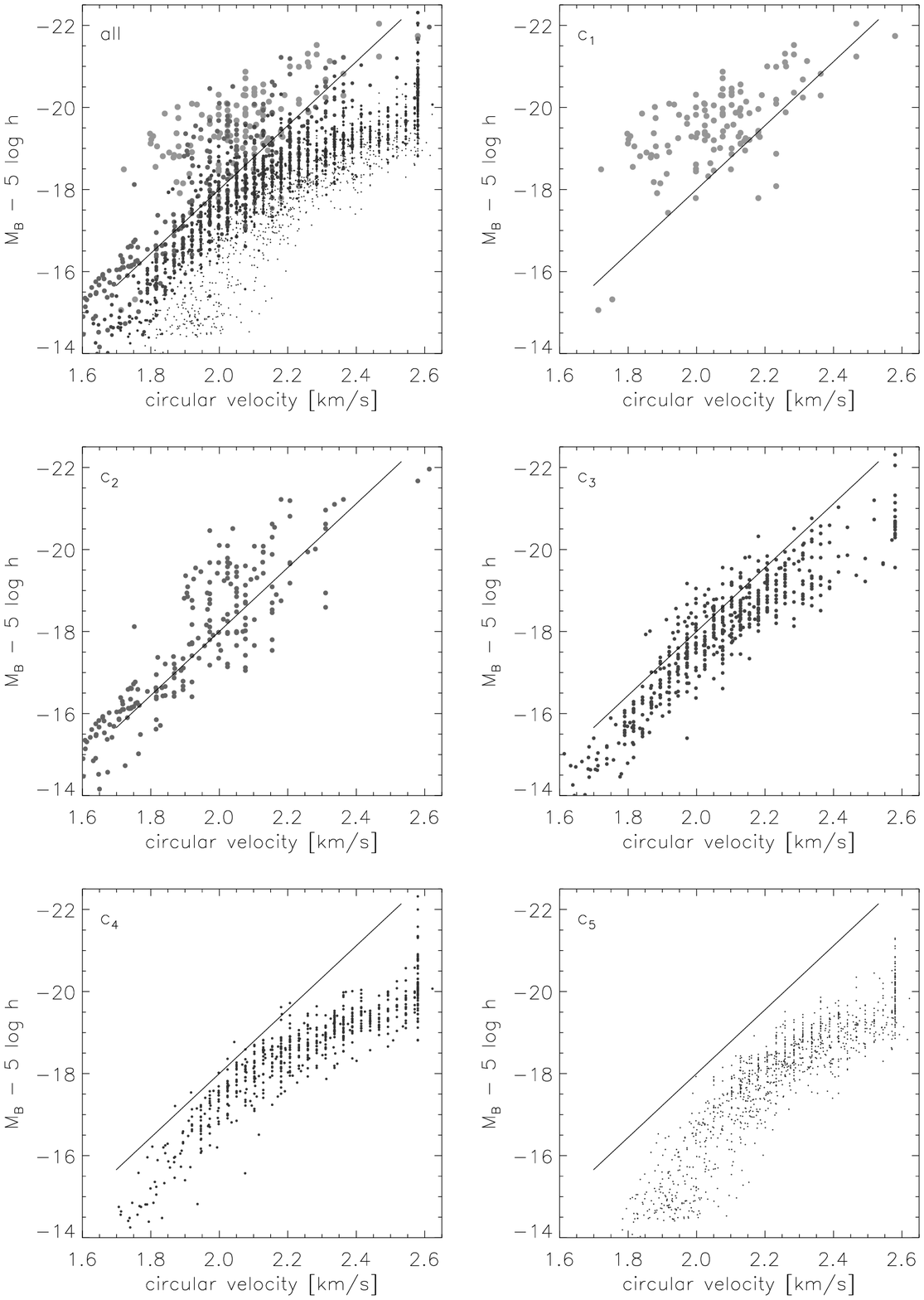,width=9truecm,height=12.5truecm}}
\caption{Luminosity-velocity (Tully-Fisher relation)
diagram for the model galaxies.
The line shows, for reference, 
the observed B-band Tully-Fisher relation for local
galaxies (Pierce \& Tully 1992). 
\label{fig:tf}}
\end{figure}

Figure~\ref{fig:tf} shows the absolute B-magnitude as a function of
circular velocity. This is essentially what is usually known as the
Tully-Fisher (TF) relation, although it should be noted that our model
magnitudes contain the effects of dust extinction, unlike observed TF
samples where at least the effects of inclination are generally
removed. Nor have we made any cut on morphology or gas fraction in the
models. This is why the slope and scatter of the relation plotted look
so different from the usual TF relation. It was shown in SP that when
the above effects are accounted for, we obtain reasonable agreement
with the observed zero-point, slope, and scatter of the I-band TF
relation in these models. The interesting thing to note is the way the
classes cut the two-dimensional space of this diagram. Galaxies in
$\c_1$ lie at preferentially {\em bright} magnitudes for their
velocity/mass (i.e., they are starbursts), whereas galaxies in $\c_5$
lie at preferentially {\em faint} magnitudes for their
velocity/mass. An increasing curvature of the relation is also seen
from $\c_1$ to $\c_5$. The progressive offset of the TF relation with
varying Hubble type is recognized (Burstein et al. 1997) but not very
well understood. This result offers a hint as to its origin, and also
suggests that the familiar observed TF relation and its small scatter
may be a special feature of the particular type of galaxies that are
generally selected for these samples.

\section{Comparison with PCA} 

\begin{figure}
\centerline{
\epsfig{file=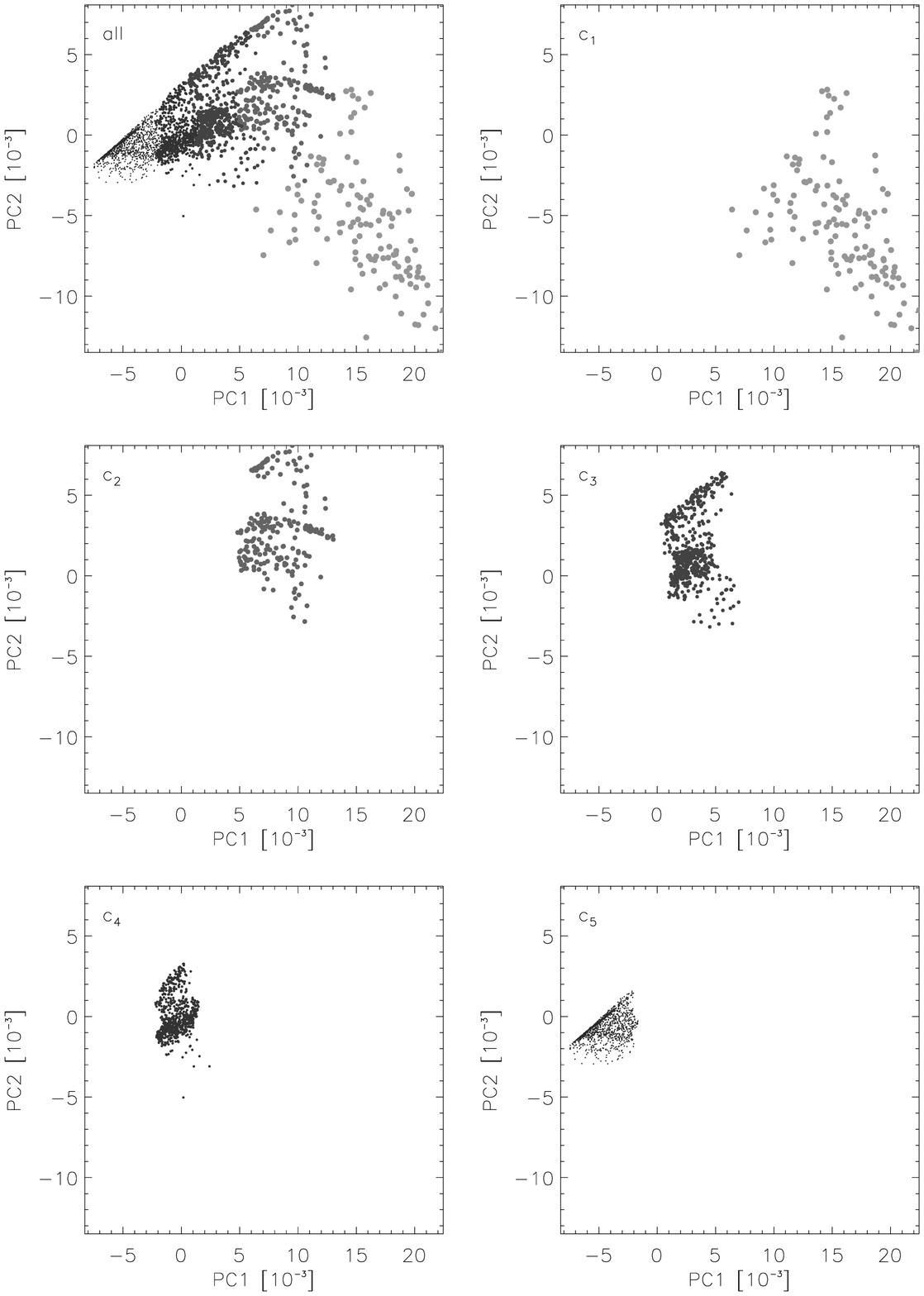,width=9truecm,height=12.5truecm}}
\caption{PCA eigenvalues $(PC1,PC2)$ of the model galaxies. 
The top left panel shows all galaxies, and the remaining panels show
separately the galaxies in each of the five IB classes.
\label{fig:pca}}
\end{figure}

PCA has previously been applied to data compression and classification
of spectral data of stars (e.g. Murtagh \& Heck 1987; Bailer-Jones et
al. 1997), QSO (e.g. Francis et al. 1992) and galaxies (e.g. Connolly
et al. 1995a; Folkes, Lahav \& Maddox 1996; Sodre \& Cuevas 1997; Galaz
\& de Lapparent 1997; Bromley et al. 1998; Glazebrook, Offer \&
Deeley 1998; Ronen et al. 1999; Folkes et al. 1999).  As we noted
earlier, while PCA operates as an efficient data compression
algorithm, it is purely linear, based only on the variance of the
distribution.  PCA on its own does not provide a rule for how to
divide the galaxies into classes.
\footnote{
For example, in Folkes et al. (1999) the classification was done by
drawing lines in the $PC1-PC2$ plane using training sets.  One
training set was based on visual inspection of the spectra by a human
expert. This led to classification which is more sensitive to emission
and absorption lines, rather than to the continuum (which is affected by
observational problems).}  It is therefore interesting to see where
the IB classes reside in the space of the PC components. The PC
eigenvectors are defined in the usual way (see the above references
for more details), and the projections of the (noiseless) model
spectra onto the first two components of this basis (PC1 and PC2) are
shown in Figure~\ref{fig:pca}.  The overall pattern reminds the one
seen in the models of Ronen et al. (1999) and in the 2dF data (Folkes
et al. 1999), but note that the PC eigenvectors are different for
every data set.  We note that our 5 IB classes form fairly
well-separated ``clumps'' in PC1-PC2 space, and that to a first
approximation, the IB classification is along PC1.  This highlights
the need to correct properly the artifacts due to optics so that the
continuum (colour) and hence PC1 can be determined accurately.  The
PC-space of the IB clumps looks quite different from the partitioning
(based on training sets) given in Folkes et al. (1999).  It has been
shown (Ronen et al. 1999) that PC1 and PC2 are correlated with colour
and emission line strength, and the sequence from $\c_1$-$\c_5$ is
again sensible in this context.

\section{Discussion}
\label{sec:conclusions}
Unsupervised classification methods are generally used to obtain
efficient representation of complex data. One can identify two general
classes of techniques for achieving this goal, geometrical and
statistical. Geometrical methods begin by an embedding of the data in
a high dimensional, usually Euclidean, space, and then searching for a
low dimensional manifold that captures the essential variation of the
data. The simplest of such methods is PCA, which provide a {\em
  linear} projection of the data. 
PCA can be generalized to more powerful linear projections,
e.g. projection pursuit (Friedman and Tukey 1974) or
to nonlinear projections that maximize statistical independence, such as
Independent Component Analysis (ICA; Bell and Sejnowski 1995).
These methods provide a low dimensional representation, or
compression, in which one might hope to
identify the relevant structure more easily.
Another approach of identifying  classes  of objects in a parameter-space
(based on a training set) is by utilising Artificial Neural Networks
(e.g. used for morphological classification of galaxies; 
Naim et al. 1995b; Lahav et al. 1996).

Statistical methods assume that the data is sampled from an underlying
distribution with some knowledge of its parametric structure. Finding
the structure of data amounts then to the estimation of the unknown
parameters of the distribution. Such methods include the familiar
Gaussian mixture estimation 
and similar vector quantization 
techniques. Modern
formulation of many of the statistical as well as geometrical methods
is based on information maximization principles for finding either
efficient compression or statistical independence among the features.

Other compression methods have also been proposed for our problem,
e.g. by starting with standard Maximum Likelihood desired parameters
(e.g. age) and utilising the Fisher information matrix to define a set
of optimal axes (Heavens, Jimenez \& Lahav 2000).

The information bottleneck method presented here provides a new
statistical approach to structure extraction. Unlike all other
techniques it aims directly at the problem of the extraction of the
{\em relevant} structure or features, where the relevance is
determined through the information on another, carefully chosen,
variable. The goal of the method is well defined and objective, with
natural information theoretical figures of merit. It is superior to
both geometric and statistical methods since it makes no
model-dependent assumptions on the data origin, nor about the
similarity or metric among data points.

An important issue, common to most unsupervised classification
methods, is model order estimation: what is the correct number of
classes? This question is closely related to the sampling noise issue
--- the obtained classes should not be sensitive to the specific
sample, thus should be robust to sampling noise. This criterion can be
checked, using techniques such as cross-validation, in most clustering
algorithms including ours. Yet it is important to emphasize that the
``true'' or ``correct'' number of classes may be an ill-defined
quantity for real data sets. The number should be determined by the
desired resolution, or preserved information in our case.

We have shown how the IB algorithm can be used to classify galaxy
spectra in a principled and objective way.  The number of distinct
classes is specified by an acceptable degree of information loss. We
have applied the algorithm to a subset of spectra from the ongoing 2dF
redshift survey, and to a mock-2dF catalogue of synthetic spectra
obtained from semi-analytic hierarchical (CDM) models of galaxy
formation.

We find that five classes preserve about 50 percent of the information
about the ensemble of 2dF spectra. The same number of classes
preserves 85 percent of the information about the model spectra in the
absence of Poisson noise. When noise is added to the models with S/N
comparable to the 2dF data, five classes preserve 75 percent of the
information. 

Examining the mean spectra of the five classes produced by our
algorithm, we first see that there is a good matching between the five
average spectra obtained for the mock data and the five average
spectra obtained for 2dF.  It is also apparent that these spectra form
a sequence from blue galaxies with strong emission lines to red
galaxies with strong absorption lines and no emission lines. This
corresponds well with the general approach usually followed in more
subjective spectral classification methods (i.e. ``by eye'').

For the model galaxies, we also show that the classes form sequences
in several physical quantities, such as the present-to-past-averaged
star formation rate (Kennicutt $b$ parameter), morphology (as
represented by the ratio of bulge-to-total stellar mass), and stellar
mass/velocity dispersion. Since the spectra obviously do not contain
any of this information directly, the existence of these correlations
(which are in accord with known observational correlations among the
physical parameters studied) seems to indicate two things. First, that
the physics used to create the model galaxies is fairly
sensible. Second, and more novel, grouping the galaxies in a way that
formally preserves the information with respect to the spectra (as our
method does), discovers interesting physical correlations.

Again for the models, we find that the classes occupy different parts
of bivariate diagrams in pairs of the physical parameters, such as age
vs. metallicity, color-color, and luminosity vs. circular velocity
(Tully-Fisher). These results may hint at important clues as to how to
constrain the star formation histories of different types of galaxies
and the physical origin of these sorts of relationships.

We compare our results with those of a Principal Component
Analysis. We find that the classes produced by the IB algorithm form
fairly well-defined clumps in the PC1-PC2 space (the projections onto
the first two principle component eigenvectors). 

We conclude that this method provides a way to classify galaxies that
is fully automated and objective, yet is related to the physical
properties of galaxies and the intuition that astronomers have built
up over the years using more subjective methods. A further advantage
is that this method can be applied in exactly the same way to
observations and models such as the ones investigated here, allowing
comparisons between theory and observations to be made on the same
footing. We intend to apply the algorithm to the full set of galaxy
spectra obtained from the 2dF redshift survey. Obvious applications
are then to study the spatial clustering of different classes and
relative biasing among them, and luminosity functions divided by
class. If spectra with adequate signal-to-noise can be obtained, the
same method could be applied to high redshift spectra to study the
evolution with redshift of different types of galaxies.

\section{acknowledgment}
We thank S. Folkes,  D. Madgwick,  A. Naim, S. Ronen, 
and the 2dFGRS team for their contribution to the work presented here.
This research was supported by a grant from the Ministry of Science, Israel.

\protect
\end{document}